\documentclass[%
 reprint,
superscriptaddress,
showpacs,
nofootinbib,
 amsmath,amssymb,aps,
prd,
floatfix,
]{revtex4-1}

\usepackage{lipsum} 

\usepackage{bm}
\usepackage{comment}
\usepackage{floatrow} 
\pdfoutput=1
\usepackage{amsmath,amsfonts,amsthm}
\usepackage{esdiff}  
\usepackage{booktabs}  
\usepackage{url}  
\usepackage{hyperref}  
\usepackage{relsize}

\usepackage{cleveref}  
	\crefname{equation}{equation}{equations}
	\crefname{figure}{figure}{figures}	
	\crefname{table}{table}{tables}
\usepackage[caption=false]{subfig}

\usepackage[normalem]{ulem}

\usepackage[usenames,dvipsnames,svgnames,table]{xcolor}

\usepackage[export]{adjustbox}
\usepackage{graphicx}

\usepackage[sc]{mathpazo} 
\usepackage[T1]{fontenc} 
\usepackage{microtype} 

\usepackage{booktabs} 
\usepackage{float} 
\usepackage{hyperref} 

\usepackage{lettrine} 
\usepackage{paralist} 

\usepackage{titlesec} 
\renewcommand\thesection{\Roman{section}} 
\renewcommand\thesubsection{\Alph{subsection}} 
\titleformat{\section}[block]{\large\scshape\centering\bfseries}{\thesection.}{1em}{} 

\titleformat{\subsection}[block]{\scshape\centering}{\thesubsection.}{1em}{} 

\DeclareCaptionFormat{myformat}{#1#2#3\hrulefill}
\usepackage{float}
\floatstyle{plaintop}
\restylefloat{table}

\DeclareMathOperator{\Tr}{Tr}

\begin{document}

\title{Inference offers a metric to constrain dynamical models of neutrino flavor transformation} %

\author{Eve Armstrong}
\email{evearmstrong.physics@gmail.com}
\affiliation{Department of Physics, New York Institute of Technology, New York, NY 10023, USA}
\affiliation{Department of Astrophysics, American Museum of Natural History, New York, NY 10024, USA}
\author{Amol V. Patwardhan}
\email{apatwardhan2@berkeley.edu}
\affiliation{Department of Physics, University of California, Berkeley, CA 94720, USA}
\affiliation{Institute for Nuclear Theory, University of Washington, Seattle, WA 98115, USA}
\author{Ermal Rrapaj}
\email{ermalrrapaj@gmail.com}
\affiliation{Department of Physics, University of California, Berkeley, CA 94720, USA}
\affiliation{School of Physics and Astronomy, University of Minnesota, Minneapolis, MN 55455, USA}
\author{Sina Fallah Ardizi}
\email{sardizi@nyit.edu}
\affiliation{Department of Physics, New York Institute of Technology, New York, NY 10023, USA}
\author{George M. Fuller}
\email{gfuller@ucsd.edu}
\affiliation{Department of Physics, University of California, San Diego, La Jolla, CA 92093-0319, USA}
\affiliation{Center for Astrophysics and Space Sciences, University of California, San Diego, La Jolla, CA 92093-0424, USA}
\date{\today}

\begin{abstract}
The multi-messenger astrophysics of compact objects presents a vast range of environments where neutrino flavor transformation may occur and may be important for nucleosynthesis, dynamics, and a detected neutrino signal. Development of efficient techniques for surveying flavor evolution solution spaces in these diverse environments, which augment and complement existing sophisticated computational tools, could leverage progress in this field.  To this end we continue our exploration of statistical data assimilation (SDA) to identify solutions to a small-scale model of neutrino flavor transformation.  SDA is a machine learning (ML) formula wherein a dynamical model is assumed to generate any measured quantities.  Specifically, we use an optimization formulation of SDA wherein a cost function is extremized via the variational method. Regions of state space in which the extremization identifies the global minimum of the cost function will correspond to parameter regimes in which a model solution can exist.  Our example study seeks to infer the flavor transformation histories of two mono-energetic neutrino beams coherently interacting with each other and with a matter background.  We require that the solution be consistent with measured neutrino flavor fluxes at the point of detection, and with constraints placed upon the flavor content at various locations along their trajectories, such as the point of emission, and the locations of the Mikheyev-Smirnov-Wolfenstein (MSW) resonances.  We show how the procedure efficiently identifies solution regimes and rules out regimes where solutions are infeasible.  Overall, results intimate the promise of this \lq\lq variational annealing\rq\rq\ methodology to efficiently probe an array of fundamental questions that traditional numerical simulation codes render difficult to access.
\end{abstract}

\maketitle 
\maketitle 
\section{INTRODUCTION}

The physics of the evolution of flavor (electron,  muon, tau) in the neutrino fields in core collapse supernovae and in neutron star binary mergers comprises an unsolved problem at the heart of the ongoing revolution in multi-messenger astrophysics. The stakes could be high. For example, finding convincing solutions to this problem might leverage gravitational wave and electromagnetic observations of binary neutron merger-generated kilonovae into deeper insights into the transport of energy, entropy, and lepton number in these sites. In turn, this could aid in understanding heavy element nucleosynthesis, connecting this problem to the issues of chemical evolution and mass assembly histories of galaxies and dark matter.

Producing self-consistent solutions, however, for the neutrino flavor evolution problem has {been} a vexing undertaking.  In part this is because of the fierce non-linearity engendered by the high neutrino fluxes and densities in these compact object environments. In short, the potentials and scattering-induced quantum de-coherence processes that govern neutrino flavor transformation themselves depend on the flavor states of the neutrinos. This non-linearity has necessitated the development of highly sophisticated numerical approaches, such as the multi-angle Neutrino BULB code~\cite{Duan:2006an, Duan:2008eb}, or the IsotropicSQA code~\cite{2019PhRvD..99l3014R}.

As with many non-linear many-body problems, initial forays into finding self-consistent numerical solutions have led to surprises. Notably, despite the small measured neutrino mass-squared differences and the large matter densities in these venues, large scale coherent collective neutrino flavor oscillations have been shown to arise, for example, in Refs.~\cite{Duan:2006an, Duan:2005cp, Duan:2006jv, Hannestad:2006nj, Cherry:2013mv, Zaizen:2019ufj, Abbar:2018shq} (see also: reviews in Refs.~\cite{Duan:2009cd, Duan:2010bg, Mirizzi:2015eza, Chakraborty:2016yeg} and references therein).  The large-scale computational tools run on supercomputers, however, are ill-suited to searching the range of relevant geometries and conditions in these environments, and their complexity sometimes leads to an obscuring of the underlying physical nature of these collective phenomena. As a result, a bevy of parallel efforts has ensued over the last 15-20 years to build smaller-scale analytic or semi-analytic models in an attempt to uncover this underlying physics. These studies~\cite{Raffelt:2007cb, Raffelt:2007xt, Dasgupta:2007ws, Johns:2017oky, 2011PhRvD..84e3013B, Raffelt:2013rqa, Mirizzi:2013wda, Duan:2014gfa, 2015PhLB..751...43A, 2016JCAP...01..028C, 2009PhRvD..79j5003S, 2016PhRvL.116h1101S, 2015PhRvD..92l5030D,  Izaguirre:2016gsx, Capozzi:2017gqd, 2018PhRvD..97b3017D, 2019PhLB..790..545A, 2019ApJ...883...80S, 2019PhRvL.122i1101C, 2019PhRvD..99j3011D, 2019ApJ...886..139N, 2019ApJ...886..139N, Johns:2019izj, Chakraborty:2019wxe, Morinaga:2018aug, Cherry:2019vkv} have revealed that the neutrino flavor field in compact objects may fall victim to a host of instabilities. {In fact, some of these} instabilities were artificially hidden from initial numerical simulations by the fact that those numerical approaches necessarily were forced to adopt, and enforce, a simplistic geometry with a high degree symmetry.

In order to utilize the understanding gained from these smaller-scale models, and to apply it to efficiently probe a range of physical conditions and parameter regimes where such physics can arise, we expand our examination of an inference-based strategy to study nonlinear neutrino flavor evolution. This work is an extension of our initial exploration described in Ref.~\cite{armstrong2017optimization}.
Specifically, we bring statistical data assimilation (SDA) to bear upon a simple neutrino flavor transformation problem.  

SDA is an inverse formulation~\cite{tarantola2005inverse}: a machine learning approach designed to optimally combine a model with data, to estimate the state of the system to an accuracy higher than that which either the data or model alone would yield.  Invented for numerical weather prediction~\cite{kimura2002numerical,kalnay2003atmospheric,evensen2009data,betts2010practical,whartenby2013number,an2017estimating} and since applied to biological neuron models~\cite{schiff2009kalman,toth2011dynamical,kostuk2012dynamical,hamilton2013real,meliza2014estimating,nogaret2016automatic,armstrong2020statistical}, SDA offers a systematic means to identify the measurements required to estimate unknown parameters of a dynamical model.  Within astrophysics, inference has been used mainly for pattern recognition~\cite{Djorgovski2006}, and its utility for model completion is gaining traction in the exoplanet community~\cite{madhusudhan2018atmospheric} and solar physics~\cite{Kitiashvili2008}. 

Further, an optimization formulation of SDA, wherein a cost function is extremized, has the ability to efficiently survey parameter space, identifying regions in which a model solution is possible and ruling out others.  This feature may lend the technique to adeptly identify physical regimes of interest, which may then be examined in more detail with the existing methodologies outlined above.

In this paper, we expand our examination of the simple steady-state model described in Ref~\cite{armstrong2017optimization}, wherein two mono-energetic neutrino beams coherently interact with each other and with a background medium, and a measurement of flavor is made at the final endpoint.  To this model we add, in a step-wise fashion, additional constraints on the neutrino flavor at the source and near the locations of the Mikheyev-Smirnov-Wolfenstein (MSW) resonance of each beam. {Such constraints on neutrino flavor along their trajectories may be motivated by, for instance, a broad physical understanding of the dynamics of neutrino decoupling, or considerations from shock reheating or heavy-element nucleosynthesis in these environments}. We seek to identify the parameter regimes that yield a solution consistent with the measurements, the constraints, and the model equations {governing flavor evolution}.  Importantly, we retain the feature of that model that it can be solved via numerical integration.  This feature offered a consistency check for SDA solutions, which is vital for the initial exploration of SDA as a viable alternative strategy for problems that numerical integration cannot probe.  In this paper, we retain that feature in order to identify the specific constraints that are required to eliminate the degeneracy of solutions that were found in the original publication.

Now, a nonlinear model will present a non-convex cost function.  With the aim to gradually freeze out a  global minimum, we employ an annealing procedure defined in terms of the rigidity of the model constraint imposed upon the calculation.  We shall demonstrate that the evolving value of the cost function over the course of annealing offers a tool for interpreting the significance of results: it reveals whether a solution has been found in the region of parameter space that has been searched. 

Ultimately, even though the system examined here is relatively simple, we intend to eventually adopt this procedure to systems with larger numbers of neutrinos and/or fewer symmetries.  This raises questions regarding scalability, since optimization-based methods tend to be computationally expensive for systems with a large number of degrees of freedom.  We comment on this aspect in Sec.~\ref{sec:discuss}.

 The manuscript proceeds as follows. In Sec.~\ref{sec:model}, we describe our physical model in terms of a system of ordinary differential equations (ODEs) governing neutrino flavor evolution. In Sec.~\ref{sec:method} lies an outline of how the model dynamics, as well as information from measurements and constraints, are incorporated into the statistical data-assimilation framework. Sec.~\ref{sec:expts} details the design of the specific experiments conducted on our physical system using the SDA framework. In Sec.~\ref{sec:result}, we summarize the results of these experiments. A discussion regarding future directions follows in Sec.~\ref{sec:discuss}.

\section{MODEL} \label{sec:model}

\subsection{\textbf{Formulation}}
The model we employ has been described in Ref.~\cite{armstrong2017optimization}, and we refer the reader there for details.  Here we describe the model's important features and the equations of motion {(flavor evolution)} used in the SDA procedure.

Two important features of the model merit comment.  First, the model is nonlinear - a key aspect of the physics that gives rise to collective neutrino flavor evolution.  SDA is particularly effective for estimating model evolution and parameter values in nonlinear models where only a subset of the state variables can be accessed experimentally.  Second, the model is sufficiently simple to be solvable via traditional forward-integration techniques. This feature enables a consistency check for SDA solutions.

We consider a two-flavor scenario in which two monoenergetic neutrino beams with different energies interact with each other and with a background consisting of particles carrying weak charge, such as nuclei, free nucleons, and electrons. The densities of the background particles and of the neutrino beams themselves are taken to dilute as some functions of a position coordinate $r$, which could be interpreted, for instance, as the distance from the neutrino sphere in a supernova.  

We write the equations of motion in terms of the \lq\lq polarization vectors\rq\rq\ $\vec P_i$ of each neutrino, after decomposing the density matrices and Hamiltonians, respectively, into bases of Pauli spin matrices\footnote{The polarization vectors (also known as Bloch vectors) are defined in terms of the neutrino density matrices: $\rho_i = \frac12(\mathbb{1}+\vec\sigma \cdot \vec P_i$). The Hamiltonian can be similarly decomposed as $H_i = \frac12(\Tr(H_i) + \vec\sigma \cdot \vec V_i)$, where $\vec V_i$ contains contributions from vacuum oscillations, neutrino-matter interactions, and neutrino-neutrino interactions, as shown in Eq.~\ref{eq:model}.} (for details see Ref.~\cite{Raffelt1993,Sigl1993}):
\begin{equation} \label{eq:model}
  \diff{\vec{P}_{i}}{r} = \left(\Delta_i \vec{B} + V(r) \hat{z} 
  +\mu(r) \sum_{j\neq i} \vec{P}_j \right) \times \vec{P}_i
\end{equation}

Here, $\Delta_{i} = \delta m^2/(2E_{i})$ are the vacuum oscillation frequencies of the two neutrinos, with energies $E_1$ and $E_2$, where $\delta m^2$ are the mass-squared differences in vacuum. $\vec{B}=\sin(2 \theta) \hat{x} -\cos(2 \theta) \hat{z}$ is the unit vector representing neutrino flavor mixing in vacuum. The functions $V(r)$ and $\mu(r)$ are the potentials for neutrino-matter and neutrino-neutrino coupling, respectively. In our model, we take the neutrino-neutrino coupling, as a function of position $r$, to be $\mu(r) = Q/r^4$. This choice is consistent with how the coupling strength varies in the neutrino bulb model calculations employing the single-angle approximation. In our SDA experiments, $Q$ is taken to be a constant with a known value.  

In contrast, the matter potential $V(r)$ is assumed to be dependent upon one or more unknown parameters, and is therefore the focus of the parameter estimation study in this paper.  In one set of experiments, the matter potential takes the form $V(r) = C_m/r^3$, where the SDA procedure is tasked with inferring the value of $C_m$, within a specified set of bounds.  In a second set of experiments, the matter potential is taken to have a slightly more complex form $V(r) = C_m(r)/r^3$, with 
\begin{equation}
C_m(r) = -f(r + L)^2 + \xi. \label{eq:Cm_var}
\end{equation}

Here, $f$ is taken to be a known constant, whereas $L$ and $\xi$ are parameters to be inferred by the SDA procedure. For further details, see \textit{Experiments}  (Sec.~\ref{sec:expts}).  All other model parameters are taken to be constant and known throughout the SDA procedure (Table~\ref{table:Known}).

\setlength{\tabcolsep}{5pt}
\begin{table}[H]
\small
\centering
\begin{tabular}{|l |c | c |c|} \toprule
\hline
 \textit{Parameter} & \textit{Value} & \textit{Initial condition} & \textit{Value} \\\midrule \hline
 $\Delta_1$ & 1000 &  $P_{1, z}(r=0)$ & 1.0\\ 
 $\Delta_2$ & 2500 & $P_{2, z}(r=0)$ & 1.0 \\
 Q & 100.0 & & \\
 $\theta$ & 0.1 & & \\\bottomrule \hline
\end{tabular}
\caption{\textbf{Model parameters taken to be known and fixed during the estimation procedure.}  The $\Delta_i$ are the vacuum oscillation frequencies of the neutrinos, and $Q$ is the multiplicative factor governing the neutrino-neutrino coupling potential $\mu(r)$.  Parameter $\theta$ is the mixing angle in vacuum.} 
\label{table:Known}
\end{table}

\subsubsection{\textbf{Physics of the model}}

 In this paper, we choose a model that is sufficiently simple from a computational standpoint, but nevertheless retains a key feature of the collective neutrino oscillation problems: nonlinearity. As a first step, we assume two neutrino beams of different energies are emitted from the source (for example, the \lq\lq neutrino sphere\rq\rq\ of a proto-neutron star) as electron flavor eigenstates.  In a core-collapse supernova environment, while all three flavors  of neutrinos and anti-neutrinos are produced in comparable quantities at late times, the neutrino flux during the early shock breakout or \lq\lq neutronization burst\rq\rq\ phase is expected to be dominated by electron neutrinos over all other flavors of neutrinos and anti-neutrinos; thus the choice of the aforementioned initial conditions was made for further simplification of the problem. On their journey through the supernova envelope, the neutrinos interact coherently with each other and with the dense ejecta surrounding the star immediately after core collapse. 

The $\hat{z}$ component of the neutrino polarization vector denotes the net flavor content of the electron flavor minus the \lq\lq $x$\rq\rq\ flavor, the latter of which can be thought of as a superposition of muon and tau flavors. Assuming  flavor evolution to be entirely forward-scattering driven, the polarization vectors are normalized to preserve particle number. At some unspecified distance, the electron and effective neutrino densities can produce a forward scattering potential that corresponds to a neutrino effective mass level crossing, which we will henceforth refer to as simply an \lq\lq MSW resonance.\rq\rq\ Large flavor conversion probability in the channel $e$ $\leftrightarrow$ $x$ may accompany MSW resonances. 

By having access only to detector measurements at the final point of our thought experiments, we can seek the optimal values of the matter density profile parameters consistent with observations, which in turn would allow us to calculate the possible location of the resonance. Additionally, one might attempt to guess the location of the resonance and use the SDA procedure to determine the existence of any matter profiles consistent with such a guess.  Another avenue for research is the consideration of the initial flavor content as a free parameter to optimize, given detector observations and an educated guess for the matter density profile; this endeavour we plan to pursue in future work.

\section{METHOD} \label{sec:method}

\subsection{\textbf{General formulation}}

Statistical data assimilation (SDA) is an inference procedure in which a dynamical system is assumed to underlie any measured quantities.  This model $\bm{F}$ can be written as a set of $D$ ordinary differential equations that evolve in some parameterization $r$ as:
\begin{align} \label{eq:ODE}
  \diff{x_a(r)}{r} &= F_a(\bm{x}(r),\bm{p}(r)); \hspace{1em} a =1,2,\ldots,D,
\end{align}
where the components $x_a$ of the vector $\bm{x}$ are the model state variables.  Unknown parameters to be estimated are contained in $\bm{p}$.  In this paper, the parameters $p$ are constant on any path in state space, although generally they may be taken to vary with position $r$.

A subset $L$ of the $D$ state variables is associated with measured quantities and constraints.  One seeks to estimate the $p$ unknown parameters and the evolution of all state variables that is consistent with the measurements and constraints provided, in order to predict model evolution at parameterized locations where the constraints are not present.

A prerequisite for estimation using real experimental data is the design of simulated experiments, where the true values of parameters are known. In addition to providing a consistency check, simulated experiments offer the opportunity to ascertain \textit{which} and \textit{how few} experimental measurements and constraints, in principle, are sufficient to complete a model.  

\subsection{\textbf{Optimization framework}} \label{subsec:framework}

SDA can be formulated as an optimization, wherein a cost function is extremized.  We take this approach, and we search the cost function via the variational method.  Importantly, we write the cost function so that the rigidity of imposed model dynamics can be adjusted.  It will be shown below in this Section that treating the \lq\lq model error\rq\rq\ as finite offers a systematic method to identify a global minimum in a specific region of state-and-parameter space.  The procedure in its entirety - that is: a variational approach to minimization coupled with an annealing method to identify a global minimum - is referred to loosely in the literature as variational annealing (VA).  The cost function $A_0$ is written in three terms: $A_0 = R_f A_{model} + R_m A_{meas} + R_{c} A_{unitarity}$.  The complete expression is shown in Equation~\ref{eq:action}, and the meanings of each term are follows: 
\begin{itemize}
 \item $A_{model}$ imposes the model evolution of all $D$ state variables $x_a$, as described by Eq.~(\ref{eq:ODE})---or more specifically in our case, by Eq.~(\ref{eq:model}).  Here, the outer sum on $n$ is taken over discretized odd-numbered grid points of the model equations of motion.  The sum on $a$ is taken over all $D$ state variables.
 \item $A_{meas}$ governs the transfer of information from the measurements and constraints $y_l$ to model states $x_l$.  It derives from the concept of mutual information of probability theory~\cite{abarbanel2013predicting}.  Here, the summation on $j$ runs over all discretized timepoints $J$ at which measurements are made, which may be some subset of all integrated timepoints of the model.  The summation on $l$ is taken over all $L$ measured quantities. 
  \item $A_{unitarity}$ represents additional  requirements imposed on variational annealing to enforce unitarity; that is, to ensure the state space solutions remain physical.  The optimization formulation we employ treats state variables as independent quantities, which is not the case for the polarization vectors describing the neutrinos.  Thus, we enforced their inter-dependence via the functions $g_i$, taken here to be: $g_i(\bm{x}(n)) = P_{i,x}^2 + P_{i,y}^2 + P_{i,z}^2$.
\end{itemize}

\begin{widetext}
\begin{equation} \label{eq:action}
\begin{split}
A_0 =& R_f A_{model} + R_m A_{meas} + R_{c} A_{unitarity}\\
A_{model}=&\frac{1}{{N}D}	\mathlarger{\sum}_{n \in \{\text{odd}\}}^{N-2} \mathlarger{\sum}_{a=1}^D \left[ \left\{x_a(n+2) - x_a(n) - \frac{\delta r}{6} [F_a(\bm{x}(n), \bm{p}) + 4F_a(\bm{x}(n+1),\bm{p}) + F_a(\bm{x}(n+2),\bm{p})]\right\}^2  \right. \\
  & \hspace{100pt} + \left.\left\{ x_a(n+1) - \frac12 \left(x_a(n)+x_a(n+2)\right) - \frac{\delta r}{8} [F_a(\bm{x}(n),\bm{p}) - F_a(\bm{x}(n+2),\bm{p})]\right\}^2 \right] \\
  A_{meas}=& \frac{1}{N_\text{meas}} \mathlarger{\sum}_j \mathlarger{\sum}_{l=1}^L (y_l(j) - x_l(j))^2\\
  A_{unitarity}=& \mathlarger{\sum}_n^{N}|g_1(\bm{x}(n))-1|^2 + \mathlarger{\sum}_n^{N}|g_2(\bm{x}(n))-1|^2.
\end{split}
\end{equation}
\end{widetext}
\noindent
$R_m$ and $R_f$ are inverse covariance matrices for the measurement and model errors, respectively.  In this paper the measurements are taken to be mutually independent, rendering these matrices diagonal.  For our purposes, $R_m$ and $R_f$ are relative weighting terms; their utility will be described immediately below in this Section.  The Lagrange multiplier $R_c$ is set to 1, and $\delta r$ is defined to be $r(n+2) - r(n)$; that is, twice the grid spacing.
One seeks the path {$\bm{X}^0 = \{\bm{x}(0),...,\bm{x}(N),\bm{p}\}$} in state space on which $A_0$ attains a minimum value.  It may interest the reader that one can derive this cost function by considering the classical physical action on a path in a state space, where the path of least action corresponds to the correct solution~\cite{abarbanel2013predicting}.  Hereafter we shall refer to the cost function of Equation~\ref{eq:action} as the action.

The procedure searches a $(D \,(N+1)+ p)$-dimensional state space, where $D$ is the number of state variables of a model, $N$ is the number of discretized steps, and $p$ is the number of unknown parameters. In the set of experiments with constant $C_m$, the number of state variables is six, one for each of the components of the polarization vectors for the two neutrino beam system. In the other set of experiments we consider $C_m(r)$ to be a state variable in addition to the polarization vectors, with equation of motion $\frac{d C_m(r)}{dr} = -\frac{f(r + L)^3}{3}$. This leads to an additional contribution to the model term in Equation~\ref{eq:action}. To perform simulated experiments, the equations of motion are integrated forward to yield simulated data (and simulated constraints), and the VA procedure is challenged to infer the parameters and the complete evolution of all state variables that were used to generate those data (and constraints).  For further details, we refer the reader to our previous publication~\cite{armstrong2017optimization}.

\subsection{\textbf{Annealing to identify a lowest minimum of the cost function}}

Our model is nonlinear, and thus the action surface will be non-convex.  The complete VA procedure anneals in terms of the ratio of model and measurement error, to gradually freeze out a global minimum of the action~\cite{ye2015systematic}.  This iteration works as follows.

We first define the coefficient of measurement error $R_m$ to be 1.0, and write the coefficient of model error $R_f$ as: $R_f = R_{f,0}\alpha^{\beta}$, where $R_{f,0} = 10^{-3}$, $\alpha = 2$, and $\beta$ is initialized at zero.  Parameter $\beta$ is the annealing parameter.  For the case in which $\beta = 0$, relatively free from model constraints the action surface is smooth and there exists one minimum of the variational problem that is consistent with the measurements.  We obtain an estimate of that minimum.  Then we increase the weight of the model term slightly, via an integer increment in $\beta$, and recalculate the action.  We do this recursively, toward the deterministic limit of $R_f \gg R_m$ (or: toward $\beta = 30$, in increments of 1.0).  The aim is to remain sufficiently near to the global minimum to not become trapped in a local minimum as the surface becomes resolved.

\section{THE EXPERIMENTS} \label{sec:expts}

\subsection{\textit{The experimental designs}}
We designed a task for the SDA procedure wherein there exist one or more unknown model parameters to be estimated in the matter potential.  The form of the matter potential is of keen theoretical interest, and it may impart a signature upon a detection.  We performed the procedure for two distinct forms of the matter potential, which will be described below in this Section.  In these experiments, the information provided to the procedure as \lq\lq measurement\rq\rq\ was the flavor of each neutrino at the endpoint of the evolution, generated using the simulated experiments.  Within the context of this simple model, by \lq\lq measurement\rq\rq\ we mean the value of $P_z$ with no noise added\footnote{We experimented briefly with the effects of noise in the measurements of $P_z$; see \textit{Results} (Sec.~\ref{sec:result}).} (of course, a real detector will measure an energy spectrum convolved with sources of contamination; see Sec.~\ref{sec:discuss} - \textit{Discussion}).

For each of the two forms for the matter potential, we designed five experiments, each defined by theoretical constraints placed upon the flavors of both neutrinos at specific locations prior to detection.  These five sets of locations were, respectively: i) at the neutrino-sphere, or the radius of emission ($r=0$); ii) at the neutrino-sphere and the center of the MSW resonances (at the radius where $P_z = 0.0$); iii) at the neutrino-sphere and near the starts and ends of the MSW resonances (at the radii where $P_z = +/- 0.9$); iv) at the neutrino-sphere and throughout the MSW resonances (at the radii where $P_z = [-0.9 : 0.9]$); v) at all discretized model locations.  In each case, we seek a solution that is consistent with both measurements and constraints.

\begin{figure}[htb] 
  \includegraphics[width=0.9\textwidth]{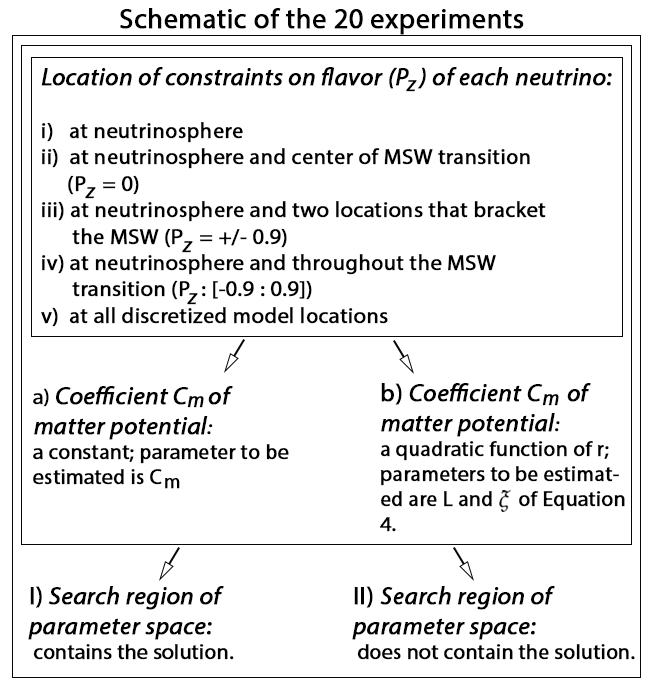}
  \caption{\textbf{Schematic of the 20 simulated DA experiments performed.}  \textit{Top}: Five distinct experiments were performed, each defined by the locations of constraints placed on flavor within the supernova envelope, for both neutrinos.  \textit{Middle}: Each of those five experiments was performed twice, once each for a distinct form of the coefficient for matter potential $C_m$. \textit{Bottom}: Each of those (ten) experiments were performed twice, once within a region of parameter space in which the true solution exists, and one within a region that does not contain the true solution.}
 \label{Fig1}
\end{figure}

Finally, for each of these five experiments, we searched a region of parameter space in which the \textit{true} solution - that is, the solution corresponding to the parameter values chosen in the simulated experiments  - exists, and alternatively a region in which the true solution does not exist.  For each region, we asked the following questions:
\begin{enumerate} 
\item \textit{Does the evolution of the action over the course of annealing reveal which paths find a solution?}
\item \textit{Do the measurements and constraints contain sufficient information to guarantee that the true solution has been (or cannot be) found---that is, do the inferred parameter values match those from the simulated experiments?}  
\end{enumerate}
Specifically, we sought to ascertain whether a plot of action versus annealing parameter $\beta$ would show unambiguously: i) that the true solution is found within the region of parameter space containing the chosen parameter values, and ii) that the true solution is \textit{not} found in the region that excludes the chosen parameter values.  We posed this question for the following reason: \textit{If the action plot can indeed identify the paths corresponding to correct solutions, then this plot can serve as a litmus test for the scenario wherein we do not know the correct solution: we can simply choose the solutions corresponding to paths of least action.}  See Figure~\ref{Fig1} for a schematic of these twenty experiments.

We calibrated our model in two deliberate ways to manipulate the ability of the SDA procedure to infer the model parameters governing the function $C_m(r)$ appearing in the matter potential $C_m(r)/r^3$.  First, the measurement and the constraint on flavor placed at the neutrino-sphere were chosen to permit a range of values for the model parameters.  Thus, for the case wherein this sole information is provided to the procedure, we expected to obtain degenerate solutions.  That is, we designed an action landscape wherein multiple global minima correspond to different paths that equally-well describe the information provided at the endpoints.  Second, additional constraints placed at the locations described above (designs ii), iii), iv), and v) in Figure~\ref{Fig1}) were designed to successively narrow the permitted range of parameter values.  Thus, as we added these constraints in succession, we expected the degeneracy to dissolve.  Examining how degeneracy-breaking works in this simple example may begin to probe the theoretically-relevant question: \textit{which, and how many, constraints are necessary to fully break degeneracy, given a particular model?}  Ultimately, such a strategy may enable an efficient identification of promising theoretical avenues versus those that can be ruled out.  Namely, the former will yield solutions that are consistent with measurements, while the latter will not.

\subsection{\textbf{The parameter to be estimated: coefficient $C_m(r)$ governing the matter potential}}

As noted, we chose two distinct forms for the matter potential.  In the first set of experiments, the function $C_m(r)$ is taken to be a constant number.  Its value in the simulated experiments was chosen so that the first neutrino, $\nu_1$, by its energy, will arrive at the midpoint of the MSW resonance (that is, attain $P_z = 0.0$) at radius $r=1$; this value of $C_m$ is 983.0. The SDA procedure was tasked with inferring this value, given the measurements and constraints. {The neutrino $\nu_2$ has a lower energy (higher $\Delta$) and therefore experiences the MSW resonance at about $r \approx 0.75$, a somewhat smaller radius.\footnote{In a realistic core-collapse supernova environment, neutrinos with typical energies of $\mathcal{O}(10)$ MeV encounter MSW resonances at radii of a few 100s to a few 1000s of kms, depending on factors such as the progenitor mass and composition, and the entropy of the environment. At these radii, the neutrino-neutrino potential $\mu(r)$ is usually sub-dominant compared to the neutrino-matter potential $V(r)$, a feature that is reflected in our toy model as well. For example, for the set of experiments with $C_m(r) \propto 1/r^3$, and for our parameter choices, we have $\mu(r) \approx 100$ at $r = 1$, compared to $V(r) \approx 983$.}}

In the second set of experiments, $C_m(r)$ is now a \textit{variable} parameter; that is, a quantity that varies but where the dynamics underlying the variability can be encoded in terms of certain fixed parameters.  We chose a simple quadratic form for this variability, given by Eq.~\ref{eq:Cm_var}, where, in the simulated experiments, we chose $f = 500.0$, $\xi = 5000.0$, and $L = 1$. This formulation effects a smooth decline in the value of $C_m$ across the regions of both MSW resonances.  The number $f$ was taken to be known and fixed, and the variability was encoded rather simply in terms of \textit{two} unknown parameters to be estimated: $L$ and $\xi$.  {In this set of experiments, neutrinos $\nu_1$ and $\nu_2$ experience their respective MSW resonances at radii of $r \approx 1.5$ and $1.2$, respectively.} It is important to begin testing the ability of the SDA procedure to handle a form for $C_m(r)$ that is variable, as ultimately it will be interesting to examine a matter potential that undergoes discontinuous changes, or shocks, throughout the envelope.

As noted, for each form for $C_m(r)$ we performed all experiments twice: once within a region of parameter space that permits the true solution, and once in a region in which the true solution does not exist.  These search regions were as follows.  For $C_m = 983.0$, the search ranges in which the true solution exists and does not exist were: $[295:2950]$ and $[2000:4000]$, respectively.  For the quadratic form of $C_m$ (with unknown parameters $L$ and $\xi$), the search ranges in which the true solution ($L = 1$ and $\xi = 5000$) exists and does not exist were: $L = [0.01:2]$ and $[2:4]$, respectively; the search range for $\xi$ was uniformly $[1500:5100]$.

\subsection{\textbf{Technical details of the procedure}}

The simulated data were generated by integrating the equations of motion of Equation~\ref{eq:model} via the Python package odeINT, with its default values.  The output of each state variable was recorded at 2001 discretized locations, with a uniform step size of 0.001.  One amendment was made to the model that is not noted in Equation~\ref{eq:model}: small offsets were added to the radius $r$ in the denominators for the matter ($1/r^3$) and neutrino coupling ($1/r^4$) potentials, to avoid divide-by-zero errors.  These offsets were, respectively: 0.001 and 0.0012.  Finally, the range for radius $r$ was taken to be 0 (at the neutrino-sphere).  The final endpoint, $r=2$, was chosen to be sufficiently far out so that the potentials $V(r)$ and $\mu(r)$ become small compared to $\Delta_i$, and the neutrinos undergo essentially vacuum oscillations at that point.

To perform the optimization, we used the open-source Interior-point Optimizer (Ipopt)~\cite{wachter2009short}.  Ipopt uses a Simpson's rule method of finite differences to discretize the state space, a Newton's method to search, and a barrier method to impose user-defined bounds that are placed upon the searches.  The discretization of the state space, the calculations of the model Jacobean and Hessian matrices, and the annealing procedure were performed via an interface with Ipopt that was written in C and Python~\cite{minAone}.  All simulations were run on a 720-core, 1440-GB, 64-bit CPU cluster.

For each of the twenty experiments defined above, twenty paths were searched, beginning at randomly-generated initial conditions for parameters and state variables. 

\section{RESULT} \label{sec:result}

\subsection{\textbf{General findings}}

General results are threefold.  First, as expected, the measurements at the detector and constraint at the neutrino-sphere permit high degeneracy of solutions, in terms of the flavor trajectories through the envelope between these two endpoints.  As we add successive constraints at specific locations within the envelope, we see that degeneracy dissolve.  Providing a constraint at the midpoint of the MSW resonance for each neutrino begins to break the degeneracy.  Adding constraints continually throughout the resonance eliminates the degeneracy entirely.  

Second: once degeneracy is eliminated, we have in the plot of action($\beta$) for any given path a litmus test for whether that path corresponds to the true solution: a path corresponding to the global minimum of the action level  corresponds to the true solution.

Third, and related to the second point above: once degeneracy is eliminated, then permitting that a sufficient number of paths is searched, their action($\beta$) plots indicate whether the true solution exists within the parameter regime that was searched.  Namely: some fraction of paths converges to the floor of the action, or none do.  (Note that one must define \lq\lq none\rq\rq\ in terms of a zero-convergence rate of paths searched - and thus must search an adequate number for one's specific purposes.)

\subsection{\textbf{Action($\beta$) plot illustrates how constraints break degeneracy, given an appropriate search region of parameter space}}

\begin{figure*}[htb]
  \includegraphics[height=\dimexpr\textheight-42pt\relax]{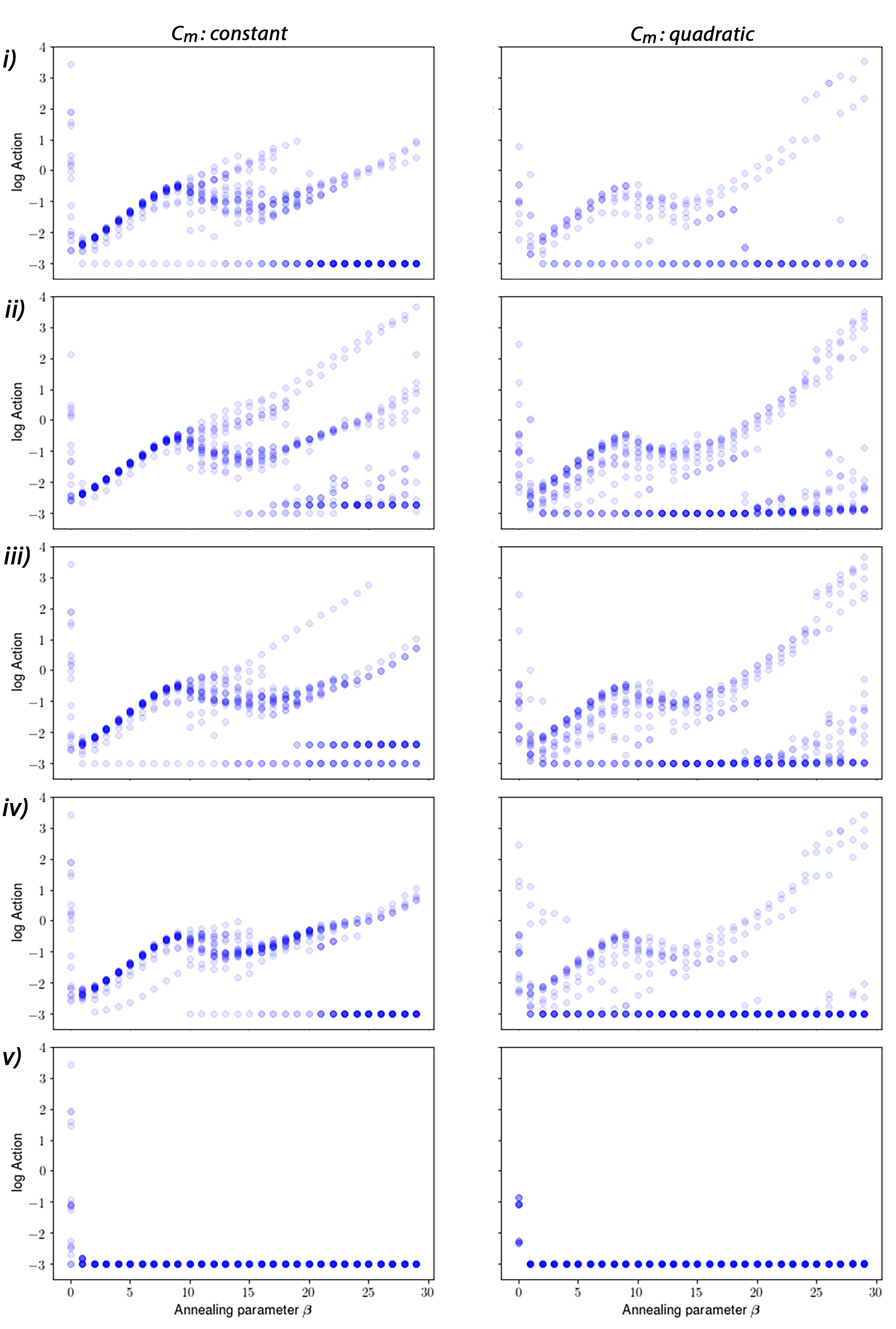} 
\caption{\textbf{Action($\beta$) for the successive constraint designs $i$ through $v$ of Figure~\ref{Fig1}}, where the solution exists within the parameter ranges that are searched.  For each experiment, all twenty paths that were searched are plotted.  \textit{Left}: $C_m =$ constant.  The true value of $C_m$ is 983.0, and the search range was: [298:2950]. \textit{right}: $C_m$ is of quadratic form, as per Eq.~(\ref{eq:Cm_var}).  For the two parameters $L$ and $\xi$ of Eq.~(\ref{eq:Cm_var}), the true value (and search range) are, respectively: 1.0 ([0.01:2]), and 5000 ([1500:5100]).}
 \label{Fig2}
\end{figure*}

Figure~\ref{Fig2} shows plots of the action as a function of annealing parameter $\beta$, where the form for $C_m$ governing the matter potential $V(r)$ is a constant number (left column) or varies as a quadratic (right column).  The five rows correspond to the five successive types of constraints imposed on flavor within the envelope (which were summarized in Figure~\ref{Fig1}).  For each row, we see a general picture: the action $A_0$ has a lowest value\footnote{This floor of the action at $10^{-3}$ is due predominantly to the additional terms added to the cost function to impose unitarity; see \textit{Unitarity terms} below in this Section.} of $10^{-3}$.  A solution corresponding to the global  minimum of the action will be consistent with both the measurements and the constraints provided, as well as with the model dynamics. 

If the sole constraint location is at the neutrino-sphere (top row), then there exist degenerate solutions corresponding to multiple global minima of the action surface. Different paths that reach a value of $A_0 = 10^{-3}$ will correspond to different solutions, each of which satisfies the measurements and constraint, and so the estimates of $P_z$ at these endpoints will be correct.  The predicted values of $P_z$, $P_x$, and $P_y$ in the interim,  however, vary across the solutions, because all describe the measured endpoint and the constraint at the neutrino-sphere equally well.  Consequently, the inferred parameter values also vary as well. {In other words, simply knowing the neutrino flavor content at the two endpoints would not be sufficient for pinpointing the location of the MSW resonance in this scenario. A range of resonance locations can be said to be consistent with a given set of endpoint constraints, as long as the strength of the matter potential is allowed to vary as a free parameter.} For an example of a state prediction that is consistent with the information provided at the endpoints but does not well match the state evolution in the interim, see  the top and bottom left panels of Figure~\ref{Fig5}.

Rows 2-5 of Figure~\ref{Fig2} then show how degeneracy is gradually broken as constraints at other locations are successively added, with regard to the flavor content near the location of the MSW resonance for each neutrino.  In the case wherein constraints are imposed at all discretized model locations throughout the resonance for each neutrino (Row 4), all degeneracy is broken, and a path that corresponds to the global  minimum of the action does indeed find the true solution\footnote{{There exist various means to quantify the \lq\lq closeness\rq\rq\ of a solution to a desired outcome, and the methodology must be chosen according to one's specific purposes.  Table~\ref{table:paramEsts} shows one example, where the mean and variance of each parameter estimate is calculated on all paths corresponding to the floor of the action, for the experiment that is found to eliminate degeneracy.}}, in terms of the predicted state and the estimated parameters.

\subsection{\textbf{Significance of the action($\beta$) plots}}

To examine the information contained in the plots of action versus annealing parameter $\beta$, let us take plots for selected individual paths.  The results in this Section are taken for the case wherein $C_m$ is a constant.  We obtained similar results for the case in which $C_m$ varies as a quadratic (not shown).

\begin{figure*}[htb]
\centering
  \includegraphics[width=0.66\textwidth]{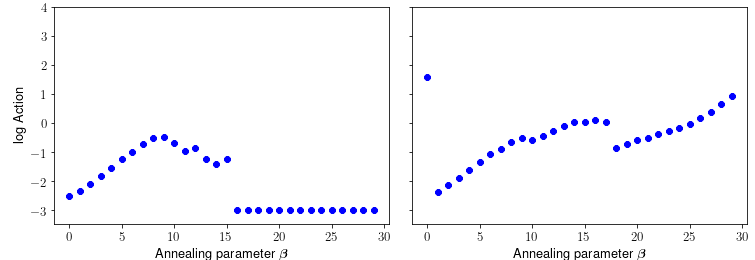}
  \caption{\textbf{Action($\beta$) on two example individual paths from Experiment $i$, wherein the sole constraint on flavor exists at the location of emission ($r = 0$)}.   The left and right plots represent 17 and 3 of the 20 total paths, respectively.  The plot at right corresponds to the state prediction shown in Figure~\ref{Fig5}, top left panel.  The constraints and measurements are obeyed, but clearly they do not unambiguously infer the state variable evolution.}
 \label{Fig3}
\end{figure*}

\begin{figure*}[htb]
\centering
  \includegraphics[width=0.99\textwidth]{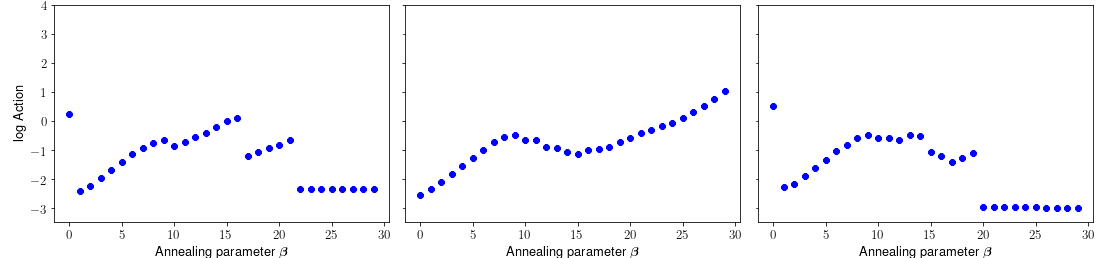}
  \caption{\textbf{Action($\beta$) on two example paths for Experiment $iii$, wherein constraints on flavor have been added at $P_z = 0.9$ and $-0.9$, for both neutrinos}.  From left to right, the three plots represent ten, five, and five paths, respectively.  {Note in the left plot the emergence of a stable \textit{local} minimum, with an action level of $\sim 10^{-2.68}$. In contrast, the plot on the right depicts a path that finds the \textit{global} minimum with action $\sim 10^{-3}$}.  The plot at right corresponds to the state prediction shown in Figure~\ref{Fig5}, bottom left panel.}
\label{Fig4}
\end{figure*}

\begin{figure*}[htb]
  \includegraphics[width=0.9\textwidth]{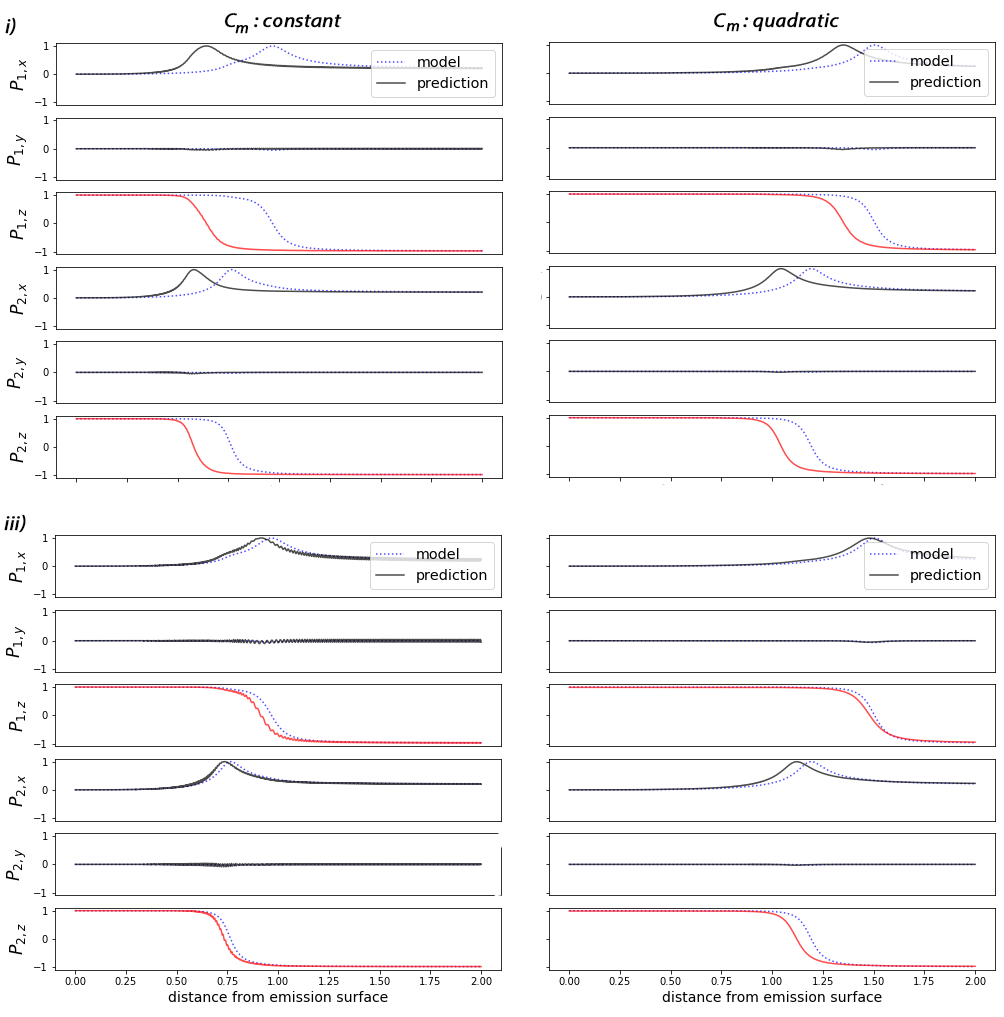}
\caption{\textbf{Representative state predictions for paths that found the lowest minimum of the action, for two sets of constraints on the flavor of both neutrinos.}  \textit{Top}: Constraints were placed only at the neutrino-sphere.  \textit{Bottom}: Additional constraints were placed at the start and end of the MSW resonance ($P_z = +/- 0.9$) for each neutrino.  \textit{Left}: For $C_m =$ constant.  The top and bottom left panels correspond to the action($\beta$) plots of Figures~\ref{Fig3} and \ref{Fig4}, respectively. \textit{Right}: For $C_m$ of quadratic form given in Eq.~(\ref{eq:Cm_var}).}
  \label{Fig5}
\end{figure*}

Figures~\ref{Fig3} and ~\ref{Fig4} shows action($\beta$) plots on selected individual paths out of the twenty that were explored for each experiment.  Figure~\ref{Fig3} shows two paths for the case where the sole constraint on flavor exists at the location of emission.  In this case, 17 of 20 paths identified the floor of the action at $10^{-3}$; that is, they appear similar to the plot at left.  This (left) plot corresponds to the best solution found, and the corresponding state prediction is shown in Figure~\ref{Fig5}, top left.  All of the seventeen cases corresponding to global minima are \lq\lq correct\rq\rq\ in the sense that they satisfy both the constraints and the model equations of motion; however, they fail to find the true solution.  Clearly, in this case the information provided solely at the endpoints of the flavor trajectories was insufficient to unambiguously infer the state variable evolution and the true values of the model parameters.

Figure~\ref{Fig4} shows action($\beta$) plots for three paths after constraints on flavor were added within the envelope at $P_z =$ +/- 0.9 for each neutrino.  Now degeneracy has begun to break.  For this experiment, five paths arrived at the action floor of $10^{-3}$, and four of those five arrived at the true  solution.  The right panel shows the action($\beta$) plot for one of these four.  The corresponding state prediction is shown in Figure~\ref{Fig5}, bottom left.

Once constraints were added at all discretized locations between $P_z = +/- 0.9$ for both neutrinos, degeneracy was eliminated. In other words, all paths that identified the floor of the action corresponded to the  true solution.  Fifteen of the twenty paths identified this solution; the mean values of the parameters across all paths corresponding to the floor of the action are shown in Table~\ref{table:paramEsts}, for the cases of both the constant and quadratic forms for $C_m$.  Note that it can be useful to \textit{begin} one's simulated experiment with this design - that is, with sufficient constraints such that degeneracy does not exist - so as to identify \textit{a priori} the optimal path and the theoretical global minimum of the action.  Then one may systematically remove constraints, to identify the minimum information required to identify a unique solution.

\setlength{\tabcolsep}{5pt}
\begin{table*}
\small
\centering
\begin{tabular}{|c c c c c | c c c c c|} \toprule
\hline
 & & \textit{$C_m =$ constant} & & & & & \textit{$C_m =$ quadratic} & &\\\midrule 
 \hline
 \textit{Parameter} & \textit{Mean} & \textit{Variance} & \textit{Search range} & \textit{True value} & \textit{Parameter} & \textit{Mean} & \textit{Variance} & \textit{Search range} & \textit{True value} \\\midrule \hline
 $C_m$ & 983.0 & $10^{-25}$ & 295.0:2950.0 & 983.0 & $L$ & 0.98 & 0.03 & 0.01: 2.0 & 1.0\\ 
 & & & & & $\xi$ & 4996.0 & 63.0 & 1500.0:5100.0 & 5000.0 \\\bottomrule
 \hline
\end{tabular}
\caption{\textbf{Parameter estimates once degeneracy is eliminated, where the search region of parameter space was known to contain the true solution.}  These numbers are taken for the case in which constraints on flavor were placed at the neutrino-sphere and at all discretized steps between $P_z = +0.9$ and $-0.9$ for each neutrino (that is, throughout the MSW resonance regions).  For each parameter, the mean value over all paths that find the floor of the action, the variance, the permitted search range, and the true parameter value are shown.  \textit{Left}: Values for $C_m =$ constant; fifteen of 20 paths found the floor of the action.  \textit{Right}: Values for $C_m$ of quadratic form; sixteen of 20 paths found the floor of the action.  These estimates are taken at a value of annealing parameter $\beta$ of 25.} 
\label{table:paramEsts}
\end{table*}

\subsection{\textbf{Action($\beta$) plot identifies relevant regions of parameter space}}

\begin{figure*}[htb]
  \includegraphics[width=0.85\textwidth,valign=t]{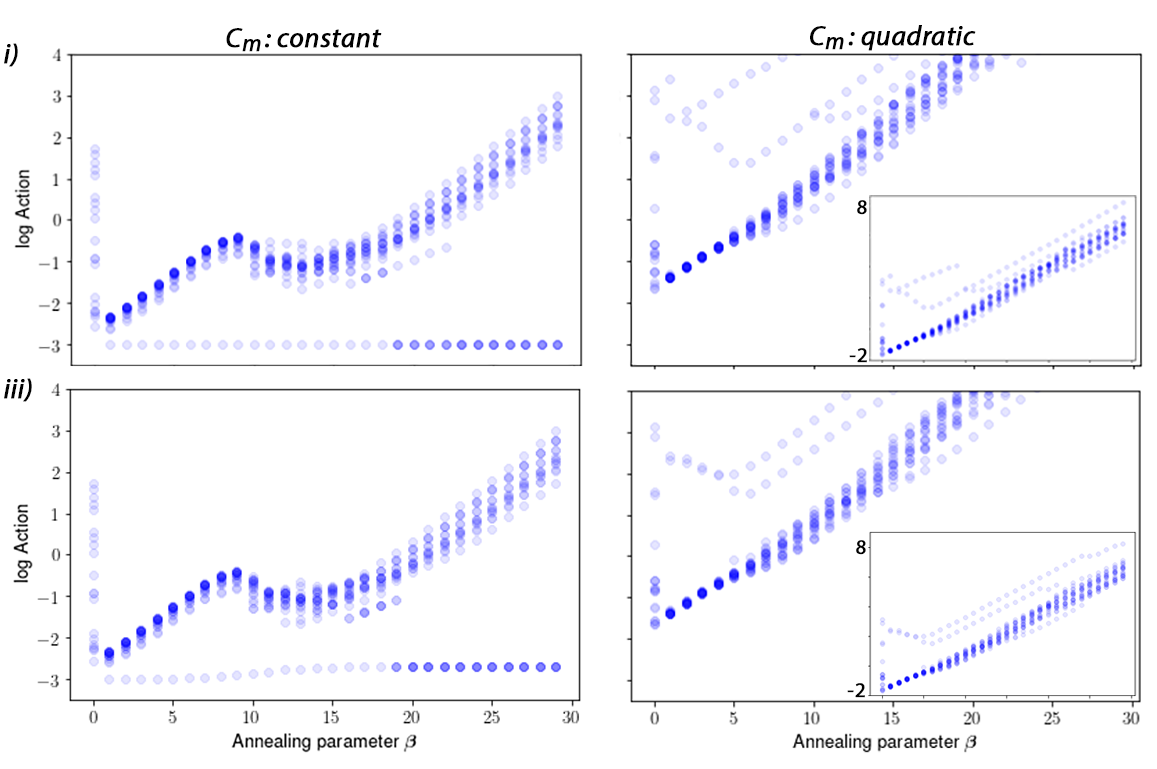}
\caption{\textbf{Action($\beta$) within a region of parameter space in which the true solution does not exist.  Once degeneracy is broken by the furnishing sufficient constraints, paths converge to a solution corresponding to an action above the lowest minimum.}  \textit{Left}: For a true value of $C_m = 983$, but using a search range of [2000:4000].  \textit{Right}: For $C_m$ of quadratic form, as per Eq.~(\ref{eq:Cm_var}) with a true value of $L=1$, but using a search range for $L$ of [2:4].  \textit{Top}: Experiment $i$, wherein a constraint is placed only at the location of emission ($r=0$).  Most paths calculate increasing model error at high $\beta$.  The few paths for $C_m=$ constant (left) that reach the lowest minimum have identified a state that is consistent with the constraints, measurements, and model, and the corresponding estimate of $C_m$ is as close as possible to the true value, given the permitted search bounds.  Clearly, there exists insufficient information to identify this region of parameter space as irrelevant.  For the case with $C_m$ of quadratic form (top right), the y-axis range cuts off the paths at high beta, and inset is their full distribution.  \textit{Bottom}: Experiment $iii$, wherein constraints are provided at the starts and ends of the resonances.  Now none of the paths find the lowest value of the action.  For the case of $C_m=$ constant (left), several paths find a value \textit{close} to the floor (at $A_0 = 10^{-2.68}$), a local minimum.  There exists sufficient information in these constraints to establish that the true solution does not exist within the parameter range that has been searched.}
 \label{Fig6}
\end{figure*}

We now demonstrate that the action plot can reveal whether one is searching a region of parameter space in which a solution consistent with measurements, constraints, and model is possible.  Figure~\ref{Fig6} shows action($\beta$) plots for experiments in which we intentionally searched a region of parameter space that excludes the true values of the matter potential coefficients, for each of the two forms of $C_m$ (constant and quadratic). We show the experiment in which constraints were provided $i)$ only at the neutrino-sphere, and  $iii)$ at the start and end of the resonances ($P_z = +0.9$ and $-0.9$).  

In the top row of Figure~\ref{Fig6}, constraints on flavor are placed at the neutrino-sphere only. In this case, for the constant $C_m$ experiment (top left), there is not sufficient information to identify this particular region of parameter space as irrelevant (not containing the true solution).  In this case, multiple paths find the lowest value for the action.  This result is as expected, as the chosen search range for $C_m$ contains values that permit an MSW resonance between $r = 0$ and $2$. Solutions consistent with the endpoint constraints of $P_z = +1$ and $-1$ can therefore be found in this case. Had the $C_m$ search range excluded values that would lead to an MSW resonance between $r = 0$ and $r = 2$, we would have expected no paths to identify the action floor.  On the contrary, for the case with $C_m$ varying as a quadratic (top right), the action increases indefinitely with $\beta$, indicating that no paths are able to find a solution consistent with the model. 
The attentive reader will notice that in this case, $C_m(r)$ is itself a state variable, and consequently the action has an additional term (as noted in Sec.~\ref{sec:method}).  In the case wherein we searched a parameter range that contained the true solution, this extra term made no numerical difference (Figure~\ref{Fig2}).  In this case, where the search range does not contain the true solution, the SDA procedure was unable to converge to a path where the model error was vanishing.

Then, with constraints added at the start and end of the resonance locations (bottom row), the procedure either settles on a stable local minimum\footnote{Note that the stable local minimum attained here may well be the lowest action level within the chosen region of the action surface; however, in this case, it does not correspond to the \textit{global} minimum, which happens to lie outside of this chosen search region.} corresponding to zero model error but nonzero measurement error (left, for constant $C_m$, wherein some paths find a value for $A_0$ of $10^{-2.68}$) or is unable to reconcile the constraints with the model, so that the model error never becomes zero and the action increases indefinitely with increasing model weight (right, for quadratic $C_m$).  As expected, these additional constraints thus help the procedure in conclusively ruling out regions of parameter space that do not contain the true solution.

{It is in this manner that inference-based methods can identify ranges of parameter space where solutions can and cannot exist. This information can then be utilized in numerical forward-integration computations to selectively explore, with greater numerical sophistication, parameter regimes where different kinds of flavor instabilities are present.}

\subsection{\textbf{Unitarity terms in the action}}

Seeking to increase computational efficiency, we repeated experiments after removing the unitarity requirement in the action formulation of Equation~\ref{eq:action}.  As noted in \textit{Method}  (Sec.~\ref{sec:method}), these constraints had been added in light of the fact that the Ipopt algorithm treats state variables as independent quantities, which is not the case for each triad of polarization vector components describing each neutrino.  These constraints are computationally demanding, and so we sought to determine whether removing them would affect solutions.  We found that, indeed, for the case of sparse constraints, the fraction of successful paths decreased by roughly an order of magnitude.  Meanwhile, the floor of the action dropped by five orders of magnitude\footnote{{ Note this drop by five orders of magnitude in the action floor - from $10^{-3}$ to $10^{-8.6}$ - upon removing the unitarity constraints in the action formulation.  This exercise exposed the dominant contribution to the action floor of $10^{-3}$.  The remaining factor of $10^{-8.6}$ is intrinsic to the structure of Ipopt and is not of scientific interest.}}, and the calculation sped by an order of magnitude.  Clearly, these unitarity terms are expensive, and yet they significantly aid the procedure in finding the solution.  In the future we will seek a compromise, for example, by annealing in the weights of these constraints.

\subsection{\textbf{Effect of additive noise}}

We repeated experiments with ten per cent noise added to the measurements.  The floor of the action rose appropriately, but the percentage of converging paths did not change.  It will be vital to consider contamination when applying this procedure to a real detection, and we plan a detailed study on the procedure's tolerance of noise in various forms; see \textit{Discussion}.

\section{DISCUSSION} \label{sec:discuss}

We have found that the value of the action over annealing will reveal whether a solution has been found within a particular region of parameter space that is consistent with the constraints, the available measurements, and the model equations of motion.  Moreover, if the provided constraints are sufficient for degeneracy breaking, the action plots can also reveal whether the \textit{true} solution - that is, the solution that reproduces the state variable evolution and the parameter values from the simulated experiments - has been found.  The efficiency with which it reveals this information suggests that the technique may well complement the numerical integration techniques referenced in \textit{Introduction} to identify regimes of interest and rule out others as irrelevant.  To this end, one might, for example, test a particular theoretical framework by  beginning in a regime wherein all degeneracy is eliminated, so as to identify the theoretical global minimum of the action \textit{a priori}.  Then constraints can be successively removed, to identify precisely which, and how few, are required to break the degeneracy, and to identify whether a solution exists within a region of interest of parameter space.

There are important issues to consider in anticipation of a real detected neutrino signal.  First, writing the model to scale is a nontrivial task: depending on the optimizing algorithm, the computational complexity increases as some power law or exponential in the number of state variables.  This issue is encountered, for example, in numerical weather prediction.  The model used at the European Centre for Medium Range Weather Forecasts contains $10^9$ state variables, out of which $10^7$ are measured~\cite{poli2013data}.  It will be instructive to draw upon the expertise in that field, as state-of-the-art forward integration codes contain $10^7$ neutrinos.  

One option may be to recast the SDA procedure as a Monte Carlo (MC) search, rather than optimization.  MC methods are more readily parallelizable, and thus may be a more realistic means to employ supercomputer clusters for large-scale simulations.  Further, MC algorithms can search a significantly larger region of state-and-parameter space, compared to a descent-only optimizer.  On the other hand, in cases where initial conditions - or the ranges of interest in parameter space - are relatively well constrained, MC algorithms can be significantly more computationally expensive than optimizers.  Moreover, the relative advantages of various algorithms for a large-scale model will depend on the details of any specific investigation.  It will be important to examine the strengths and limitations of various techniques, to identify the most feasible means of tackling a particular question regarding a large-scale model. 

A related study will be a detailed examination of the procedure's sensitivity to contamination in measurements.  Any Earth-based detection will include a signal convolved with various possible sources of contamination; it will be vital to develop a reliable method to recognize real physical signatures.  In addition, the measurement error in the cost function must allow for a transfer function between measurement and the associated state variable. 

Prior to preparing for real data, we plan to examine the procedure's performance in a host of other simulated scenarios.  We are particularly interested in bipolar oscillations and fast flavor conversions. A survey of parameter space using inference-based methods, exploring regimes where these behaviors can manifest, would be a valuable exercise.  It will be intriguing to examine what new insights the inference framework yields regarding these regimes.

\section{ACKNOWLEDGEMENTS}

We thank H.~Abarbanel for helpful conversations. E.~A. and S.~F.~A acknowledge an Institutional Support for Research and Creativity grant from New York Institute of Technology. A.~V.~P., E.~R., and G.~M.~F. acknowledge the NSF (grant no. PHY-1630782) and the Heising-Simons Foundation  (2017-228). Additionally, G.~M.~F acknowledges NSF Grant Nos. PHY-1614864 and PHY-1914242, from the Department of Energy Scientific Discovery through Advanced Computing (SciDAC-4) grant register No. SN60152 (award number de-sc0018297). Thanks also to the good people of Doylestown, Ohio.

\bibliography{bib}
\end{document}